**Tuning the Fano Resonance with an Intruder Continuum**

J. Fransson[1], M.-G. Kang[2], Y. Yoon[2], S. Xiao[2], Y. Ochiai[3], J. L. Reno[4], N. Aoki[3] & J. P. Bird[2,3,*]

1: Department of Physics and Astronomy, Uppsala University, Box 534, SE-751 21, Uppsala, Sweden

2: Department of Electrical Engineering, University at Buffalo, the State University of New York, Buffalo, NY 14260-1900, USA

3. Graduate School of Advanced Integration Science, Chiba University, 1-33 Yayoi-cho, Inage-ku, Chiba 263-8522, Japan

4. CINT/Sandia National Laboratories, Dept. 1131, MS 1303, Albuquerque, NM 87185, USA

* Corresponding author: jbird@buffalo.edu, +1-716-645-1015

Through a combination of experiment and theory we establish the possibility of achieving strong tuning of Fano resonances (FRs), by allowing their usual two-path geometry to interfere with an additional, "intruder", continuum. As the coupling strength to this intruder is varied, we predict strong modulations of the resonance lineshape that, in principle at least, may exceed the amplitude of the original FR itself. For a proof-of-concept demonstration of this phenomenon, we construct a nanoscale interferometer from nonlocally-coupled quantum point contacts and utilize the unique features of their density of states to realize the intruder. External control of the intruder coupling is enabled by means of an applied magnetic field, in the presence of which we demonstrate the predicted distortions of the FR. This general scheme for resonant control should be broadly applicable to a variety of wave-based systems, opening up the possibility of new applications in areas such as chemical and biological sensing, and secure communications.

Arising from the coupling of a discrete level to a continuum, the Fano resonance (FR) [1] is ubiquitous to atomic-, molecular-, and condensed-matter systems [2-4]. As a wave-interference phenomenon, it is also of increasing importance in optics, plasmonics, and metamaterials, where its ability to cause rapid signal modulations under variation of some suitable parameter makes it desirable for a variety of applications [5]. The observation of this resonance in plasmonic nanostructures [6-10], for example, provides a sensitive mechanism for chemical and biological sensing. In metamaterials, in contrast, FRs are being explored for use in slow-light devices, electromagnetically-induced transparency and electro-optic switching [5,11-13]. In the field of nanoelectronics, FRs have been widely observed in semiconductor quantum dots [4], and have motivated a number of proposals for electrical- or optical-signal control [4,14-22]. One possible application is in spintronics, where FRs are being explored as a means to filter spins in an initially-unpolarized current [4,21,22]. Other work [20] has demonstrated a nonlinear FR in optically-pumped quantum dots, a phenomenon that could find use as a method to detect weak couplings of two-level quantum systems.

A surprising feature of the FR is that its rich variations are obtained from a simple interference geometry. As first formulated by Fano [1], this involves an electron that is excited from an initial atomic level ($|i>$) to a final state ($|f>$) in a continuum, reaching this via two separate paths. The first of these is direct, while the second involves an intermediate transition through a discrete level. Transmission from $|i>$ to $|f>$ is determined by the phase interference of these two paths, dependent upon whose relative amplitudes FRs with various lineshapes can be obtained. A generalized representation of this scheme is presented in Fig. 1(a), in which the direct path from $|i>$ to $|f>$ has matrix element $r$, while transmission via the discrete level ($D$) is described by matrix elements $v$. (It should be noted that both Figs. 1(a) & 1(b) represent transitions between quantum states in some "configuration space", and that the various states involved may even be at the same energy.) A need that has emerged from the possible technological applications of FRs is to tune their lineshape to achieve signal modulation. In recent studies of the nonlinear FR mentioned above, this was achieved by using optical pumping to manipulate the coupling between quantum-dot states and a physically-separate continuum, thereby allowing the FR to effectively be turned "on" or "off" [20].

While the two-path Fano interferometer provides a powerful scheme for manipulating wave transmission, in this Letter we demonstrate the possibility of achieving even stronger resonant control, by making just a simple modification to the interference geometry. The essential element of our approach is to couple an additional continuum (*C*) to the original Fano system, thereby forming a three-path interferometer (Fig. 1(b)). By varying the strength of the coupling (*t*) to this continuum, we show that it is possible to induce significant lineshape distortions, which may even exceed the amplitude of the original FR. Motivated by the connection of this phenomenon to that of *q*-reversal in Rydberg atoms [23] – in which the FR due to a given atomic level may undergo dramatic phase reversal when it interacts with a separate "intruder" level – we refer to the additional continuum as an intruder. For a proof-of-concept demonstration of its influence, we realize the intruder in a nanoscale interferometer formed by a pair of coupled quantum point contacts (QPCs) [24,25]. Exploiting the unique aspects of the density of states in these structures, we configure a system analogous to that of Fig. 1(b) by forming the discrete level and the intruder within one QPC. The second is then used as a detector whose continuum transport is affected by its wavefunction overlap with the other QPC. Through suitable control of the relevant gate voltages, a FR is induced in the detector by bringing it into energetic resonance with the discrete level [26-29]. The resonance develops a pronounced and systematic distortion in a magnetic field, behavior that we show here can be attributed to the ability of this parameter to tune the coupling between the intruder and the two-path interferometer.

While our proof-of-concept demonstration of the intruder focuses on a solid-state implementation, capable only of operation at cryogenic temperatures, it must be emphasized that the general scheme presented in Fig. 1(b) should be broadly applicable to a variety of wave-based systems. This in turn should open up the possibility of new applications, including those in chemical and biological sensing [5]. Another possibility is in secure communications, where the role of the intruder could be played by an undesirable eavesdropper, whose attempts to intercept secure communications could be sensitively detected as an associated distortion of the FR.

Experiments were performed on coupled QPCs realized in epitaxially-grown GaAs/AlGaAs heterostructures (Sandia samples EA750, EA755 & VA0284). Each of these wafers had a 30-nm thick GaAs quantum well, inside of which a high-quality two-dimensional electron gas was formed. The density of this ranged from $1.8 - 2.5 \times 10^{11}$ cm$^{-2}$ at 4.2 K, with corresponding mobility of $1.4 - 3.7 \times 10^{6}$ cm$^2$/Vs. While in this Letter we focus on new results obtained from measurements of the device of Refs. 28 & 29, in the Supporting Information we present corroborating results from a second device. Measurements were performed by placing the samples in the vacuum can of a liquid-helium cryostat, and subjecting them to the magnetic field generated by a superconducting solenoid. Our investigations were performed in the temperature range of 4.2 – 40 K, where effects arising from ballistic carrier transport could be resolved.

Our experiments make use of the peculiar properties of QPCs close to pinch-off, where strongly-enhanced carrier interactions can spontaneously distort their self-consistent potential. The distortion is associated with Friedel oscillations generated by scattering from the QPC barrier, and is thought to lead to the formation of a narrow well that may support a quasi-localized state (LS) near the center of the QPC [30-35]. Several studies have been undertaken to explore the structure of this unusual microscopic feature [35-38], and we have revealed its presence by using it as the discrete level in a FR geometry [26-29]. In these experiments (see the Supporting Information for further details on the measurement procedure), two QPCs are formed in close proximity and are allowed to interact via an intervening region of two-dimensional electron gas (Fig. 2(a)). By suitable variation of the voltage ($V_s$) applied to the gates of one of the QPCs (referred to hereafter as the "swept-QPC", and denoted by the subscript "$s$") we form a LS inside it by pinching it off. With the detector-QPC (denoted hereafter by the subscript "$d$") biased at fixed gate voltage ($V_d$, typically chosen to ensure that this device operates well above pinch-off), a resonance is then observed in its conductance ($G_d$) when the variation of $V_s$ drives the LS through the Fermi level. The resonance may be viewed as arising in the process indicated in the upper panel of Fig. 2(b). Here, an electron injected from an initial state in the detector reaches a final state in the drain, either directly (matrix element $r$) or in a process in which it tunnels to and from the LS (matrix element $v$ for each process) before reaching the drain. By summing over initial and final states of this type, the result is a FR in the detector conductance [27] that corresponds well to that found in experiment.

In Figs. 2(c) & 3(a), we illustrate the evolution of the detector resonance (at 4.2 K) when a magnetic field ($B$) is applied normal to the plane of the two-dimensional electron gas. In each of the four panels of Fig. 2(c), we show the detector FR and the associated variation of the swept-QPC conductance ($G_s$). Consistent with our prior work [29], at zero field we find that the detector resonance is only weakly asymmetric. With increasing magnetic field, however, this feature undergoes a dramatic evolution, developing a pronounced dip that evolves systematically on its less-negative gate-voltage side. Importantly (see Supporting Information), the shape of this new resonance cannot be described by the universal form [1] typical of FRs. These lineshape distortions are far too strong to arise from Zeeman splitting [28], and also cannot be attributed to magneto-dispersion of the detector states. While the magnetic field does depopulate [24,25] the subbands of the detector, resulting in an overall decrease of its conductance (see Fig. S6 of the Supporting Information), our experiments are configured in a manner such that the influence of this is never strong enough to pinch the detector off. Under such operation, prior work has shown that the detector resonance remains essentially unaffected by even large changes in background conductance [28].

In Fig. 4(a), we demonstrate that the high-field resonance has an extremely unusual temperature dependence. With increase of temperature above 4.2 K its peak is quickly suppressed while the dip grows more pronounced. The overall effect is therefore of a transition to an anti-resonance, whose structure becomes more clearly resolved with *increasing* temperature.

Key to understanding the results described above is the expected form of the density of states in the swept-QPC at pinch-off. This form is sketched in the schematics of Fig. 2(b), in which we indicate two distinct components; a narrow peak corresponding to the LS and a quasi-continuous spectrum. The latter feature corresponds to the one-dimensional (1D) subbands that govern transport when the swept-QPC is open, but which should be driven above the Fermi level once it is pinched-off. Both schematics indicate the situation for which the LS is aligned with the reservoir Fermi level, and so is in resonance with the detector. At zero magnetic field (upper schematic) the 1D subbands lie well above the LS and so have little influence on this resonance. This situation changes in a magnetic field, however, due to its tendency to induce compression of the various electronic levels [39,40]. This is shown in Fig. 2(d) where we plot, for a non-interacting model, the magneto-dispersion of the LS and the edge of the lowest 1D subband. These calculations were performed by assuming that the potential of the swept-QPC at pinch-off consists of a narrow well, embedded at the saddle point of the two-dimensional barrier more normally associated with QPCs (see the Supporting Information). The LS is therefore modeled with a two-dimensional harmonic confinement ($\hbar\omega_{LS}$ = 1 meV), while the 1D subbands are attributed to the more common transverse confinement ($\hbar\omega_{1D}$ = 3 meV), allowing their magneto-dispersions to be analytically calculated [39,40]. Figure 2(d) shows a clear trend for the LS to approach the lowest 1D subband with increasing magnetic field, a result suggesting that the latter may function as an intruder and influence the FR due to the LS. As sketched In the lower schematic of Fig. 2(b), the detector resonance observed in such a situation will arise from a three-path interference phenomenon, whose direct path once again involves transmission from an initial state in the detector to a final one in the drain (path *r*). Interfering with this channel are two further pathways: one in which the electron tunnels to and from the LS (characterized by matrix elements *v*) before reaching the drain, and; a second in which the tunneling instead takes the electron to the lowest 1D subband and back. It is this last process that represents the contribution of the intruder, and the magnetic field should serve as the control parameter that allows its coupling to be varied. Note here that while this mechanism may not immediately appear to resemble that in Fig. 1(b), it does nonetheless have the essential feature of three separate paths with different intermediate steps ($|i> \to |f>$, $|i> \to D \to |f>$ & $|i> \to C \to |f>$).

For a more detailed analysis of our experiment, we have formulated (Supporting Information) a generalized theoretical description of the three-path interferometer of our experiment. In Fig. 1(c), we demonstrate the possibility of using this scheme to induce produced signal modulations, whose amplitude can even exceed that of the original FR. Specifically shown in this figure is the energy-dependent conductance ($\delta G_d$, related to the transmission) from $|i>$ to $|f>$, for various values of the intruder coupling. A general result which follows from our analysis is that the full resonance lineshape in this system may be interpreted as arising from two contributions; the first being the usual Fano mechanism, while the second involves interference of the intruder with the original continuum. Although the former process ($\delta G_{LS}$, inset to Fig. 1(c)) yields a nearly-symmetric resonance (i.e. $q \gg 0$) for the set of parameters considered here, the contribution from the intruder ($\delta G_s$, Fig. 1(d)) shows a pronounced anti-resonance. The location of this feature is determined by the energy separation assumed between the discrete level and the intruder, whose density of states is taken to be of 1D character. From a comparison of Figs. 1(c) & 1(d), it is clear that the overall resonance is strongly affected by the intruder, which distorts its line-shape dramatically; while the peak now arises from the usual Fano mechanism, the minimum is due to the two-continuum interference and the separate control of these two processes, as enabled by the coupling *t*, provides an effective means for resonant control.

In Fig. 3(b), we show how the evolution of the detector resonance in the magnetic field may be reproduced within our model, by direct variation of the intruder coupling *t*. For simplicity in these calculations, we have assumed a fixed detuning of 0.1 meV between the LS and the intruder, motivated by our expectation that the separation between these states should vary only slowly at high fields (see Fig. 2(d)). In this way, we are able to compute resonance curves that compare very well with those obtained experimentally. The implication, therefore, is that the magnetic-field induced distortion of the detector resonance does indeed result from increased coupling to an intruder, in this case one that is formed by the lowest 1D subband of the swept-QPC. In support of this scenario, we note that the most pronounced change in the detector resonance occurs between 0- and 2-T, beyond which little further change is observed (compare the 2- and 4-T plots in Fig. 2(c)). Such behavior is consistent with the results of Fig. 2(d), in which the intruder quickly approaches the LS with initial increase of the magnetic field, but thereafter advances on it much more slowly. A further point that can be made here follows by comparing the variations of $G_s$ and $G_d$ (Fig. 2(c)), which shows that the peak in the detector resonance always occurs once the swept QPC is fully pinched-off (where $G_s \sim 0$). The dip, on the other hand, occurs at less-negative $V_s$, where the QPC is beginning to open. This sequence is consistent with the form of the density of states suggested in Fig. 2(b), in which the LS lies at lower energy than the 1D subbands and so should require stronger gate biasing to align it with the Fermi level in the drain. At the same time, the correlation of the dip to the onset of $G_s$ is fully consistent with the idea that the dip occurs on achieving alignment between the edge of the 1D continuum and this Fermi level.

Turning to the temperature-dependent evolution of the detector resonance in Fig. 4(a), this behavior cannot be reproduced by adopting the usual thermal smearing of the Fermi function. Instead, in Fig. 4(b) we demonstrate that that the essential features of experiment can be captured by mimicking the increase of temperature by simultaneously decreasing the coupling ($v$) to the LS while increasing that ($t$) to the intruder. In terms of justifying this ad-hoc approach, we note that the confinement of the LS should weaken with increasing temperature, and previous experiment [28] suggests that the characteristic temperature scale on which this occurs corresponds well to that for which the peak is suppressed in Fig. 4(a). At the same time, since the intruder lies at higher energy than the LS, activation into this component should increase with increasing temperature. In the Supporting Information, we show that reasonable agreement with experiment can be obtained by varying the coupling to the LS alone. Although the quality of the resulting fits is not as good as in Fig. 4(b), this provides confidence that we capture the essential physics of the intruder. Most importantly, the essential conclusion reached from our calculations is that the unusual temperature dependence of Fig. 4(a) directly reflects the fact that the peak and dip of the high-field resonance are due to physically distinct states (LS vs. 1D intruder, respectively). Indeed, it is interesting that at 40 K, where all evidence of the peak is suppressed, the high-energy tail of the anti-resonance in Fig. 4(a) is suggestive of the 1D form of the intruder density of states.

Concluding with some general comments, while we have focused here on providing a solid-state implementation of the intruder-modified FR, it must be emphasized that the interference scheme we have proposed is a completely general one. As such, it should be broadly applicable across a variety of different wave-based systems, including those in both photonics and electronics. The intruder provides an effective means to tune the FR, and this characteristic could enable sensitive schemes for chemical or biological sensing, and for the modulation of photonic or electronic signals. In the field of secure communications, for example, the intruder could represent an undesirable eavesdropper whose attempts to intercept a secure transmission channel could be detected directly as a sensitive distortion of its FR. Perhaps most importantly, however, our study suggests the value of pursuing extended geometries for interference, in which the two-path Fano interferometer serves merely as the basic building block.

## ASSOCIATED CONTENT

**SUPPORTING INFORMATION**

**Supporting Information Available:** Detailed theoretical derivation of the influence of the intruder on the Fano resonance, as well as a simple analytical model for the magneto-dispersion of the QPC localized state and the edge of the 1D subbands. Experimental details include a description of measurement techniques, and a demonstration of the intruder in an alternative gate configuration. This material is available free of charge via the Internet at http://pubs.acs.org.

## AUTHOR INFORMATION

**Corresponding Author:** Jonathan Bird, jbird@buffalo.edu, phone: +1 (716) 645-1015

**Author Contributions:** The authors contributed equally to this work.

**Notes:** The authors declare no competing financial interest.

## ACKNOWLEDGEMENTS

The experimental research was supported by the U.S. Department of Energy, Office of Basic Energy Sciences, Division of Materials Sciences and Engineering under Award DE-FG02-04ER46180. The work was performed, in part, at the Center for Integrated Nanotechnologies, a U.S. Department of Energy, Office of Basic Energy Sciences user facility. Sandia National Laboratories is a multi-program laboratory managed and operated by Sandia Corporation, a wholly owned subsidiary of Lockheed Martin Corporation, for the U.S. Department of Energy's National Nuclear Security Administration under contract DE-AC04-94AL85000. JF acknowledges support from the Swedish Research Council.

**FIGURE CAPTIONS**

**Figure 1.** (a) Schematic illustration of the interference geometry involved in the conventional Fano resonance. (b) Modified Fano geometry with the "intruder" continuum ($C$). (c) Influence of the intruder on the Fano-resonance lineshape. The main panel shows the variation of the overall resonance lineshape ($\delta G_d$) as a function of $t$, while the inset shows the original resonance ($\delta G_{LS}$) due to the two-path geometry when $t = 0$. For the specific values of $r$ (= 0.1) and $v$ (= 4) chosen here, this resonance is only weakly asymmetric. (d) The contribution to the resonance in (c) from continuum-continuum interference alone ($\delta G_s$), for various values of $t$. In both (c) and (d), the zero of energy corresponds to that of the discrete level ($D$) and the detuning between $C$ and $D$ is held fixed at ~0.1 meV.

**Figure 2.** (a) Electron-microscope image of the coupled-QPC device studied here, showing the biasing scheme for the various gates. The upper-left pair of gates define the detector-QPC while the lower-left pair form the swept-QPC (circled). The interference pathways ($r$, $v$ & $t$) for the detector resonance are also indicated. Gates denoted "Gnd" are held at ground potential and so have no influence on the experiment. For this reason these structures have been artificially lightened in the image. (b) Schematics illustrating how the coupled-QPCs provide a realization of two- (top) and three- (bottom) path interferometers. Top and bottom panels are for $B = 0$ and $B > 0$, respectively, and the corresponding form of the density of states (DoS) in the swept-QPC is also indicated. Red denotes the LS, while the blue curve corresponds to the continuum arising from the 1D subbands. (c) Evolution of the detector resonance at 4.2 K in a perpendicular magnetic field. Blue data are the resonant contribution to the detector-QPC conductance ($\delta G_d$), while the red curves show the corresponding variation of $G_s$. $\delta G_d$ was obtained by subtracting a monotonic background from $G_d(V_s)$, as described in Ref. 29. The dotted red line in each panel denotes pinch-off of the swept-QPC, identified as the vanishing of $G_s$. (d) Calculated (non-interacting) dispersion of the LS and the bottom of the lowest 1D subband as a function of $B$. The inset shows the corresponding separation of the two levels as a function of the field, and the dotted line denotes the asymptotic variation of the level spacing. See Supporting Information for further details of the calculation.

**Figure 3.** (a) Experimental variation of the detector resonance at 4.2 K for several magnetic fields (indicated). (b) Corresponding theoretical curves, using *t* as the control parameter (values indicated). The calculations consider fixed detuning (~0.1 meV) between the LS and the intruder, and the curves have their amplitudes normalized to match experiment. Successive curves in both (a) and (b) are shifted upwards in increments of $0.025 \times 2e^2/h$.

**Figure 4.** (a) Detector resonance measured at *B* = 8 T and various temperatures (indicated). The insets highlight the behavior at the highest and lowest temperature. (b) Calculated resonance obtained by simultaneously increasing transmission via the intruder (*t* = 1.10 – 1.60) while decreasing that via the discrete level (*v* = 6.80 – 2.80). The insets highlight the behavior obtained at the two ends of this series (coupling parameters indicated in each plot).

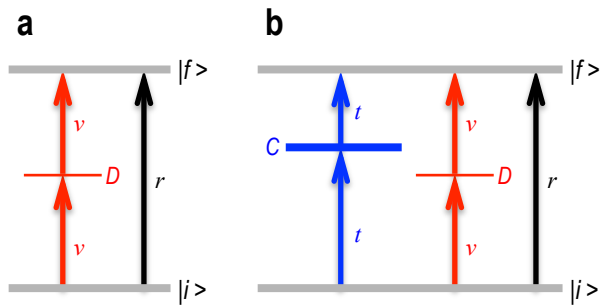
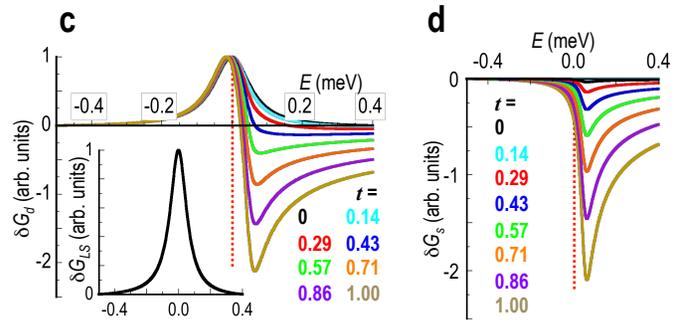

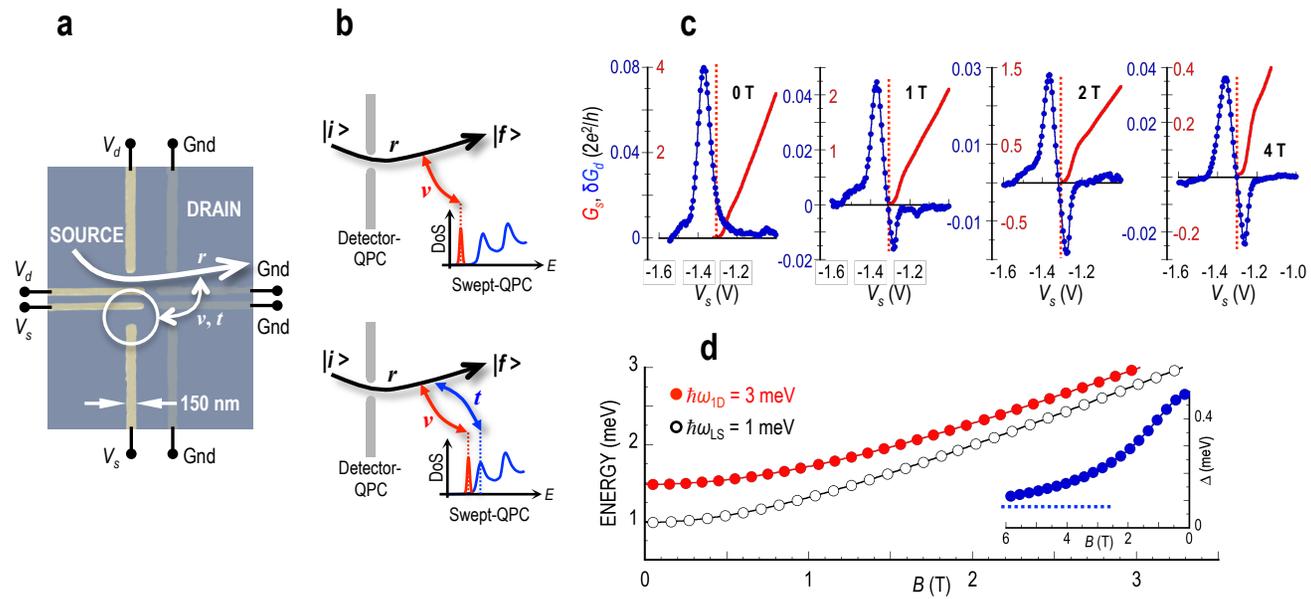

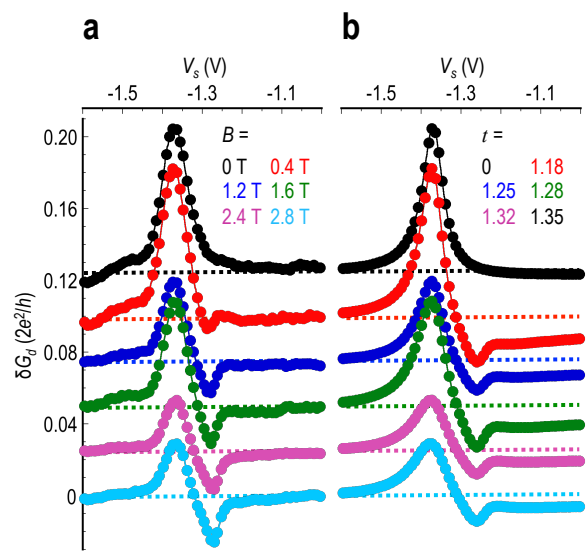

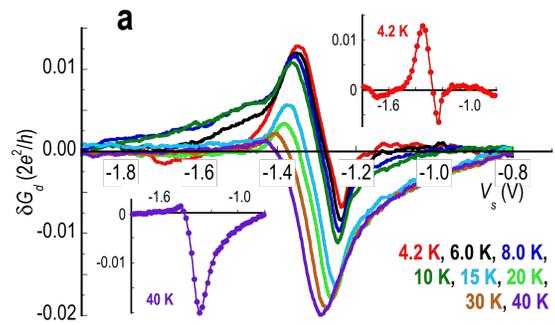

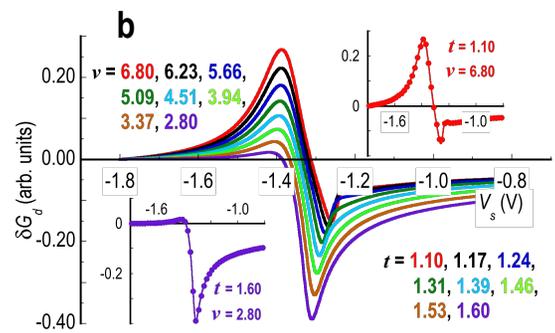

Table of Contents Graphic

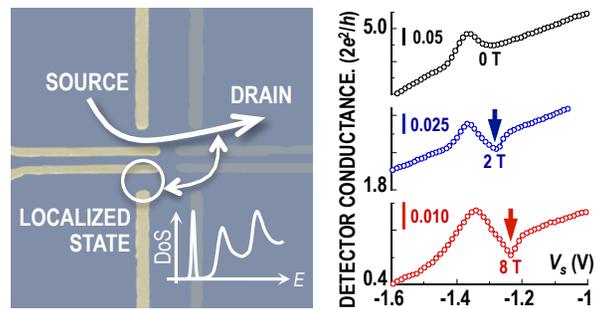